\documentclass[aps,prl,twocolumn,showpacs,superscriptaddress,groupedaddress,10pt]{revtex4-1}

\usepackage{amsmath}
\usepackage{bm}
\usepackage{graphicx}
\usepackage{color}
\usepackage{upgreek}
\usepackage[linkcolor = blue, citecolor = blue, urlcolor = blue, colorlinks = true]{hyperref}
\usepackage[capitalise]{cleveref}
\usepackage[version=3]{mhchem}
\usepackage{textcomp}
\usepackage{commath,amssymb}

\newcommand{\nm}{\, \mathrm{n m}}
\newcommand{\perml}{\, \mathrm{m l^{-1}}}
\newcommand{\pH}{\, \mathrm{p H}}
\newcommand{\uM}{\, \mathrm{\upmu M}}
\newcommand{\mM}{\, \mathrm{m M}}

\newcommand{\ugperml}{\, \mathrm{\upmu g m l^{-1}}}
\newcommand{\umsqpers}{\, \mathrm{\upmu m^2 s^{-1}}}
\newcommand{\us}{\, \mathrm{\upmu s}}
\newcommand{\umpers}{\, \mathrm{\upmu m s^{-1}}}
\newcommand{\pers}{\, \mathrm{s^{-1}}}
\newcommand{\degree}{\,^{\circ}}
\newcommand{\second}{\, \mathrm{s}}
\newcommand{\percent}{\, \mathrm{\%}}
\newcommand{\um}{\, \mathrm{\upmu m}}
\newcommand{\mm}{\, \mathrm{mm}}

\begin{document}

\title{Filling an emulsion drop with motile bacteria}

\author{I.~D.~Vladescu}
    \email{i.vladescu@ed.ac.uk}
\author{E.~J.~Marsden}
    \email{e.j.marsden@ed.ac.uk}
\author{J.~Schwarz-Linek}
\author{V.~A.~Martinez}
\author{J.~Arlt}
\author{A.~N.~Morozov}
\author{D.~Marenduzzo}
\author{M.~E.~Cates}
\author{W.~C.~K.~Poon}
    \email{w.poon@ed.ac.uk}
\affiliation{SUPA and The School of Physics \& Astronomy, The University of
Edinburgh, King's Buildings, Mayfield Road, Edinburgh EH9 3JZ, United Kingdom}
\date{\today}


\begin{abstract}
We have  measured the spatial distribution of motile \textit{Escherichia coli} inside spherical water droplets emulsified in oil. At low cell concentrations, the cell density peaks at the water-oil interface; at increasing concentration, the bulk of each droplet fills up uniformly while the surface peak remains. Simulations and theory show that the bulk density results from a `traffic' of cells leaving the surface layer, increasingly due to cell-cell scattering as the surface coverage rises above $\sim 10\%$. Our findings show similarities with the physics of a rarefied gas in a spherical cavity with attractive walls.
\end{abstract}

\pacs{Valid PACS appear here}

\maketitle

The physics of self-propelled particles~\cite{poon2013physics} -- natural or synthetic `swimmers' -- is an active area  of current condensed matter and statistical physics, where understanding non-equilibrium effects poses a `grand challenge'. Swimmers are \emph{intrinsically} out of equilibrium even without external driving. There is as yet no general recipe for predicting their collective behavior.

Confinement is of significant interest in diverse areas of physics. In this context, it is fascinating to note that self-propelled particles can confine themselves spontaneously. Motile \textit{Escherichia coli} and other bacteria encountering a surface continue to swim along it, giving rise to self-organised confinement to a 2D layer. Experimentally, the number density of motile \textit{E.~coli} between two parallel glass slides peaks strongly at the slides~\cite{li2009accumulation,berke2008hydrodynamic}. However, in this geometry, the cell density between the walls remains low and there is little 3D confinement, because the `wall-hugging' swimmers can escape essentially to infinity along the wall. Even the 2D confinement is weak: surface swimmers spread out to minimise interaction, and single-body physics suffices to explain `wall hugging'~\cite{li2009accumulation,berke2008hydrodynamic,Gompper,GoldsteinWall}.

The interesting question now arises: what would happen if there is confinement in all spatial dimensions? Biologically, motile bacteria in nature are sometimes confined in this way, e.g. in raindrops~\cite{yang2009flagellar} or infected host cells~\cite{southwick1994dynamic,southwick1994dynamic}, possibly leading to motility~loss~\cite{melilotiReview}. In physics, the collective behavior of confined swimmers has attracted recent interest. At high density, motile {\it Bacillus subtilis} in a cylindrical drop develop stable vortices~\cite{wioland2013confinement,Lushi2014}, while simulations of swimmers in a 2D box suggest novel forms of phase separation near close packing in the absence of hydrodynamic interactions~\cite{Yang}. In this work, we perform experiments starting from the opposite limit, and probe the way in which interaction effects emerge amongst motile bacteria confined within spherical emulsion droplets as the average swimmer density, $\rho_0$, increases from a small value.

As expected, at $\rho_0 \rightarrow 0$, we observe motile cells `hugging' the inner surface in a layer~\cite{li2009accumulation,berke2008hydrodynamic}. As $\rho_0$ increases, we find that the drop fills up in an unexpected way: the bulk density increases uniformly while the surface peak remains. We present simulations and theory that reproduce essential features of our observations, and which suggest that the physics is reminiscent of a rarefied gas in a spherical cavity with attractive walls~\cite{Cieplak1999}.



\begin{figure}
    \centering
\includegraphics[width = 0.9 \columnwidth]{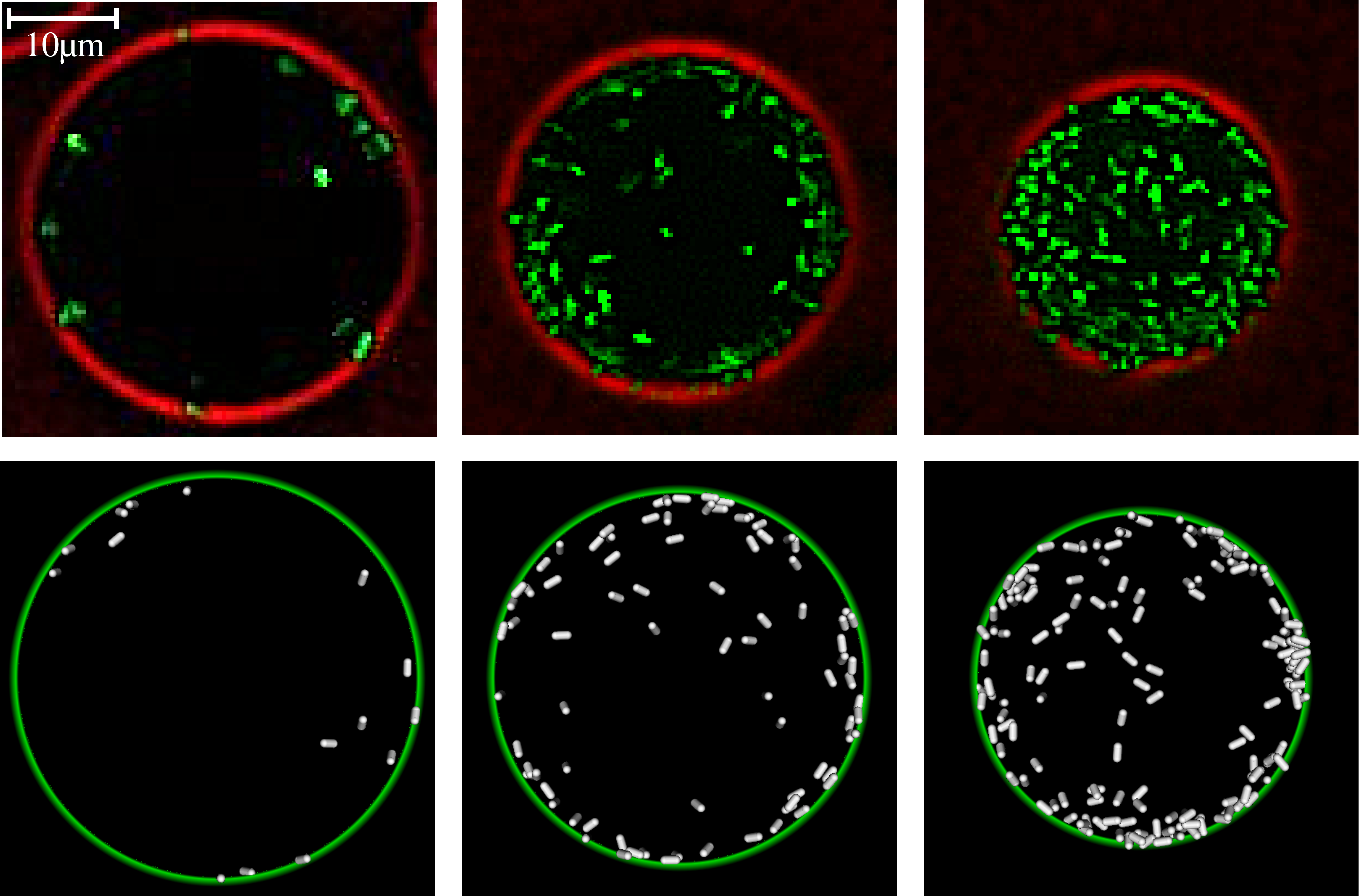}
\caption{Snapshots of $2\um$-thick cross-sections of droplets, $(\phi, R) = 0.47\%, 17.4\um$; $2.7\%, 16.1\um$; $6.2\%, 14.3\um$ (left to right). Upper: Experiment, confocal images, $\sim 2\um$ below the equator, with red droplet edges and green bacteria. Lower: Simulated equatorial cross sections, with white bacteria and green droplet rims.  }
    \label{snapshots}
\end{figure}

We studied spherical water-in-oil emulsion drops with a range of radii, $R$, containing a smooth-swimming mutant of {\it Escherichia coli} AB1157 bacteria in phosphate motility buffer at increasing average cell density, $\rho_0$. Green fluorescent protein expressed by the bacteria and a dye that preferentially adsorbs to the emulsifier on the water-drop surfaces allowed us to take high-resolution fluorescent confocal image stacks and reconstruct the cell positions inside droplets, \cref{snapshots}. The majority of droplets had $R = 10$-$20\um$ (Fig.~S1), and are therefore significantly smaller than the persistence length of our swimmers (estimated to be $\lambda \gtrsim 100\mu$m). {\it In situ} optical characterisation using an oxygen-sensitive dye~\cite{RTDP} and differential dynamic microscopy~\cite{wilson2011differential,martinez2012differential} confirmed the absence of oxygen gradients within these droplets and that cells swam during the duration of our experiments with essentially a constant speed distribution. Taking care to minimise aberration effects arising from working with spherical drops (Figs.~S2, S3), we counted cells within concentric shells to obtain the cell density as a function of distance from the center, $\rho(r)$, which we assume to be isotropic~\footnote{See supplementary material at \url{http ...} for details.} for all preparation and imaging details.)

\begin{figure}
    \centering

    \includegraphics[width = 0.49 \columnwidth]{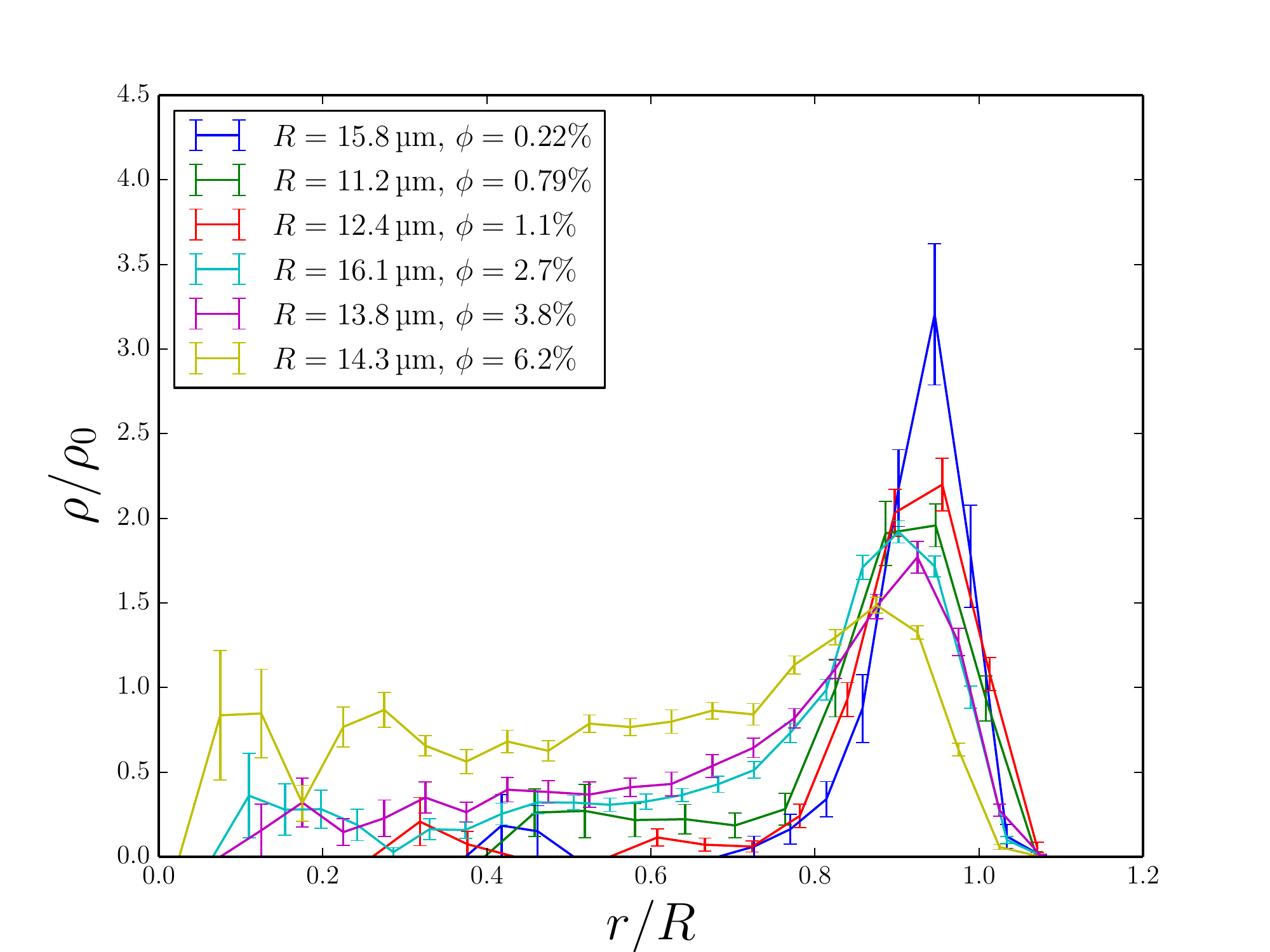}
    \includegraphics[width = 0.49 \columnwidth]{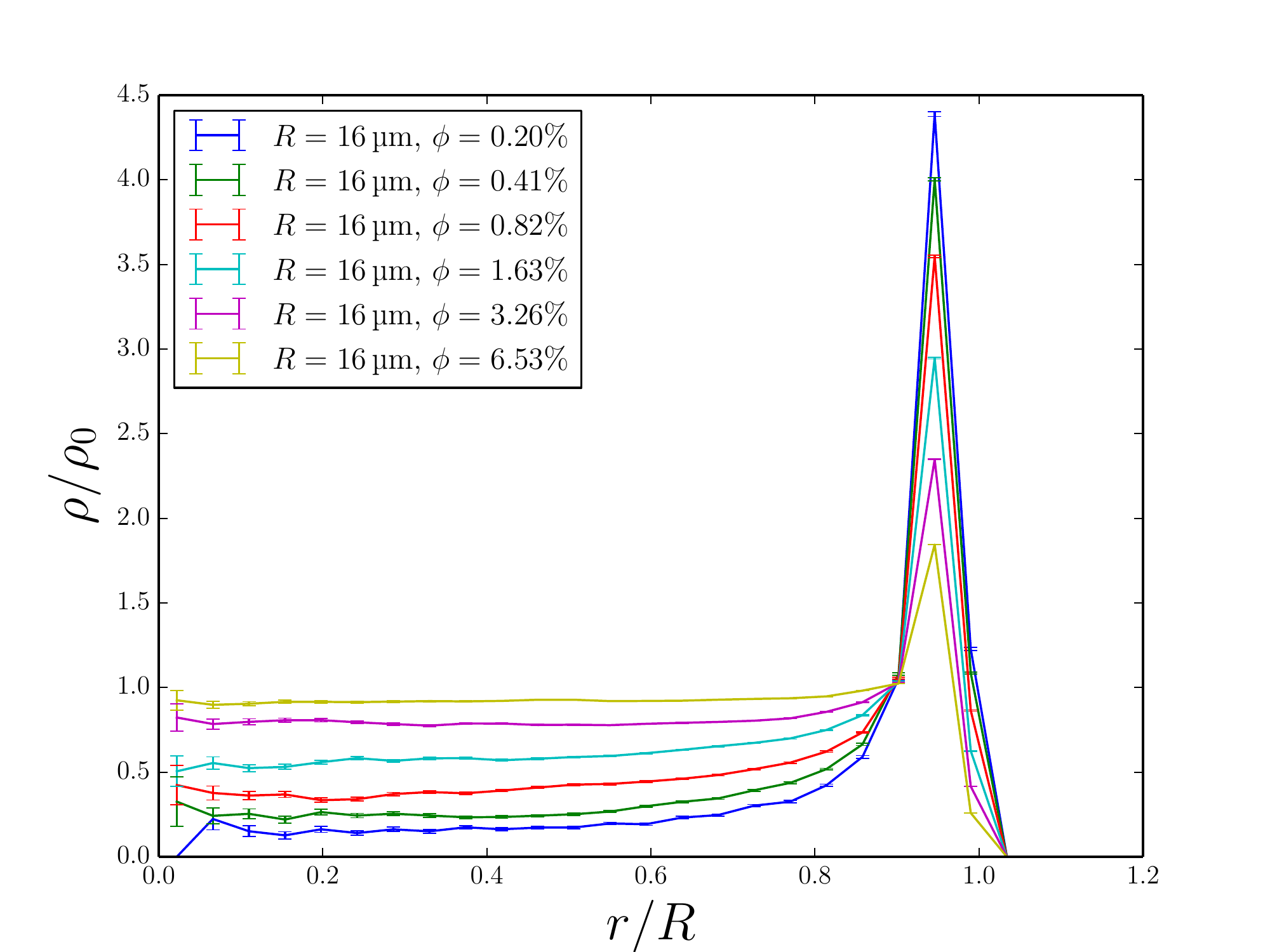}
    
    \caption{Radial bacterial number density distributions, $\rho(r)$, normalised by the average number density of the whole droplet, $\rho_0$, plotted against the radial distance from the center, $r$, normalised by the droplet radius, $R$, averaged over 10 data sets. Left: Experimental data for $R \simeq 14\um$. Right: Simulation data for $R = 16\um$ over a similar range of volume fractions as in experiment obtained for the case of maximal scattering at cell-cell collision.}

    \label{histograms}
\end{figure}

Typical density distributions, $\rho(r)$, for $R = (14 \pm 2)\um$ over a range of $\phi_0$ (calculated by taking each cell  to be a $2\um \times 1\um$ spherocylinder)  are shown in \cref{histograms}, with $\rho(r)$ normalised by the average number density $\rho_0$ and the radius by the droplet radius $R$. Each curve is the result of averaging over 10 stacks of analysed images.

At low $\phi_0$, bacteria are localized in a shell beneath the water-oil interface. Visually, almost all of these were motile, although a few non-motile cells were also localised at this interface. As $\phi_0$ increases, the peak in $\rho/\rho_0$ drops and migrates inwards, while the cell density throughout the rest of the drop increases uniformly. This scenario occurs at all $R$ studied (Fig.~S4), although there was not enough statistics to investigate $R$-dependence systematically. It is easily explained why our findings are qualitatively independent of droplet size. We work with $2 R < \lambda$; a qualitative change is only expected when droplets become larger than the swimmers' persistence length, $\lambda$.

\begin{figure}
    \centering

     \setbox1=\hbox{\includegraphics[width = 0.807 \columnwidth]{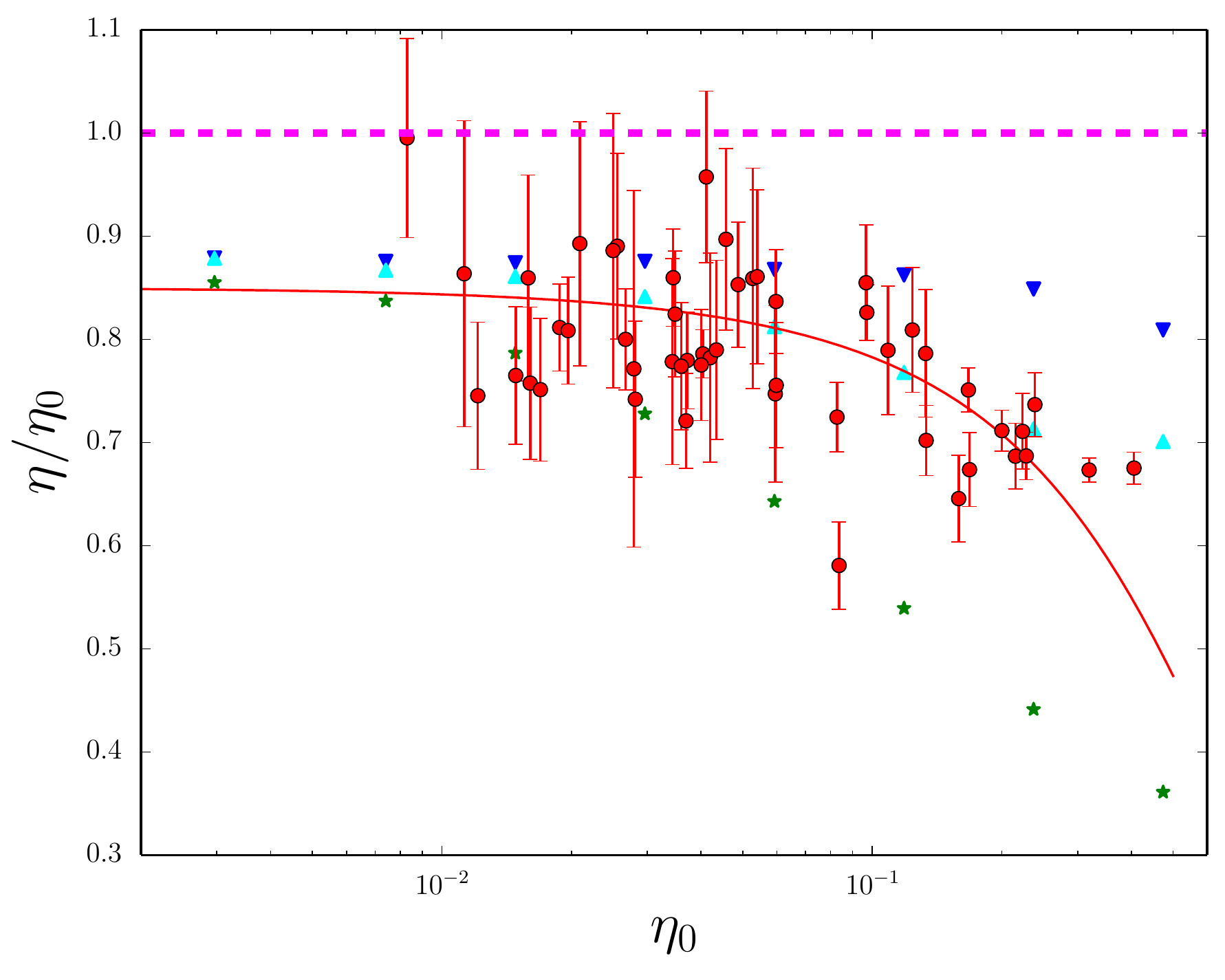}}
    \includegraphics[width = 0.9 \columnwidth]{etaf.pdf}
    \llap{\raisebox{0.8cm}{\makebox[\wd1][l]{\includegraphics[width = 0.36 \columnwidth]{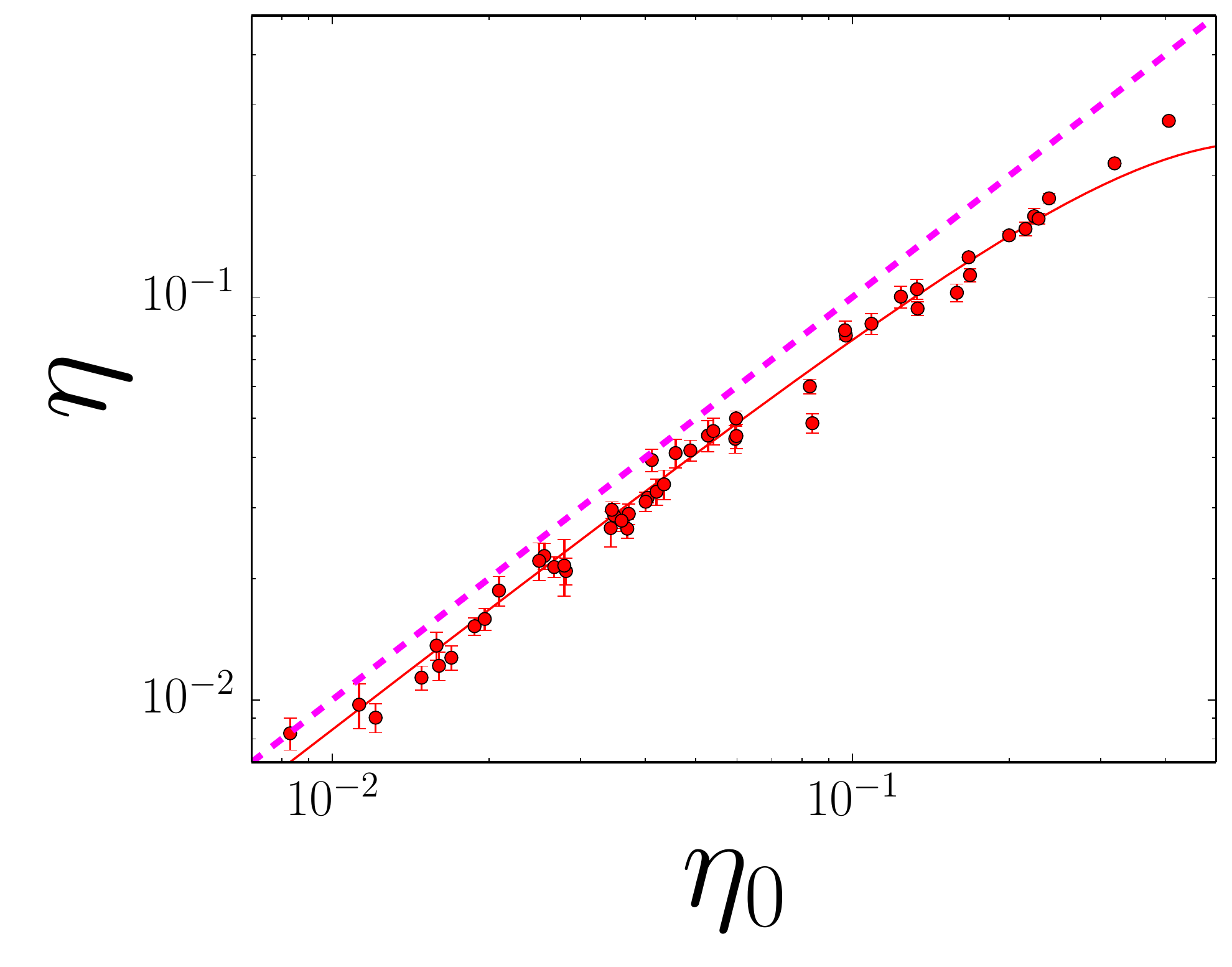}}}}
    \caption{The surface area fraction of bacteria, $\eta$, expressed as a fraction of the total potential surface coverage, $\eta_0$, as a function of $\eta_0$, from experiment ({\color{red} $\bullet$}) and simulations ($\theta_r^{(c)} = 0$ {\color{blue} $\blacktriangledown$}; $\theta_r^{(c)} \simeq 0.032$, {\color{cyan} $\blacktriangle$}; $\theta_r^{(c)} = \pi$, {\color{green} $*$}). Inset: The same data plotted in terms of the un-normalised surface coverage $\eta$: the absolute number of bacteria at the droplet surface increases sublinearly with the average cell density. In both inset and main figure, the red solid line is fit to \cref{model} and the pink dashed line represents total surface accumulation.}
    \label{eta}
\end{figure}

While the $\rho(r)/\rho_0$ peak decreases with $\phi_0$, \cref{histograms}, the absolute number of cells `hugging' the surface increases with $\phi_0$. \cref{eta} (inset) shows the total number of cells, $N_s$, found within the peak \footnote{Throughout, the peak is taken to span $R_p < r < R$, where $R_p$ is where the density first rises to $\rho = \rho_0$ coming from the centre. Other reasonable algorithmic or visual definitions of `the peak' do not change our conclusions.} plotted against the total number of cells, $N_0$. We report $N_s$ as the total area the surface cells would cover as a monolayer, $N_s A_b$ (where $A_b =$ area covered by one cell) normalised by the droplet surface area ($A = 4 \pi R^2$), $\eta = N_s A_b / A$ (the `surface area fraction'); $N_0$ is also reported as the total cell area $N A_b$ similarly normalised, $\eta_0 = N_0 A_b / A$.

Note that the peaks in \cref{histograms} are much larger than crowding-induced layering in confined hard particles. For hard spheres in a rigid spherical cavity, there is no surface peak until $\phi_0 \gtrsim 20\percent$~\cite{HScavity}; we see well-developed peaks at $\phi_0 \ll 5\percent$ due to `wall hugging'~\cite{li2009accumulation,berke2008hydrodynamic} and not crowding.

The bulk of the droplet fills in a surprising way as $\phi_0$ increases. Given wall hugging, one might expect that the first surface layer would act as a `wall' for a second layer to accumulate, etc. Such layering was indeed observed for motile \textit{B. subtilis} in a cylindrical droplet. As $\phi_0$ increases, cells build up in layers, `leaving the center almost empty' (see supplementary video 3 in \cite{wioland2013confinement}). Instead, the bulk of our spherical droplets fill up uniformly with {\it E. coli} as $\phi_0$ increases, \cref{histograms}.

A qualitative explanation is as follows. At $\phi_0 \rightarrow 0$, cells are found almost exclusively at the inner droplet surface due to wall hugging~\cite{li2009accumulation,berke2008hydrodynamic,Gompper,GoldsteinWall}. This lower-density surface layer is as yet non-interacting. As $\phi_0$ becomes finite and more and more cells arrive at the interface, the surface coverage eventually reaches a point when there will only be room for another cell if an existing surface cell leaves, spontaneously due to reorientation, or by scattering off the arriving cell or with another surface cell. Since $\lambda > 2R$ and the bulk density remains relatively low, a `departing' cell most likely travels along an approximately straight trajectory to another part of the interface, where the process repeats. This cross-droplet `traffic'  is manifested as a uniform increase in bulk density.

Consider first a simple analytic model for this picture. Bacteria with speed $v$ give rise to a uniform flux towards a flat surface. We measure the bulk and surface concentrations using the equivalent area fractions $\eta_0$ and $\eta$ already introduced, so that the inward flux is  $\propto (\eta_0 - \eta)v$,  where the factor $\eta_0 - \eta$ measures the number of cells in the bulk. We assume that an arriving cell will be trapped at the surface if it is presented with empty surface, the latter with probability $\propto (1 - \eta)$. Thus, the arriving flux $\propto (1 - \eta)(\eta_0 - \eta) v$. Surface cells also swim with speed $v$, and remain trapped until they either leave spontaneously with probability $\propto \eta$ due to orientational fluctuations, or are scattered by other cells with probability $\propto  \eta^2$. In the steady state, fluxes to and from the surface balance:
\begin{equation}
    (1 - \eta)(\eta_0 - \eta) = c \eta + b \eta^2 \,,
    \label{model}
\end{equation}
i.e., the incoming flux (LHS) is balanced (RHS) by cells leaving the surface due to spontaneous reorientation (`$c$' term) and scattering (`$b$' term). For swimmers of characteristic size $a$ confined to a region of characteristic size $R$, a simple kinematic calculation give \cite{Note1} $c \simeq R / v \tau$, where $\tau$ is a typical reorientation time after which a surface cell spontaneously returns to the bulk, and $b \simeq k R / a$, where $k$ is the probability that a surface cell-cell collision scatters a cell into the bulk.

Our experimental data, Fig.~\ref{eta} ({\color{red}$\bullet$}), though noisy at the lowest average cell densities, can nevertheless be fitted to \cref{model} (\cref{eta}, solid line, main and inset), giving $c = 0.14 \pm 0.01$ and $b = 0.7 \pm 0.1$. In the low density limit ($\eta \rightarrow 0$), we can set $b = 0$ in \cref{model} (no scattering) and find $\eta/\eta_0 = (1 + c)^{-1}  \simeq (1 - c)$, i.e. all but a fraction $c$ of cells accumulate at the surface. Clearly $c = 0$ (dashed line, Fig.~\ref{eta} main and inset) is untenable. Since $R \simeq 15\um$ and $v \simeq 10\umpers$, the fitted $c \simeq R / v \tau = 0.14$ gives $\tau^{-1} \simeq 0.1\pers$. Encouragingly, this is comparable to our estimate of the free rotational diffusion coefficient of a single flagellated cell, $\mathrm{D_r} \simeq 0.2\pers$ \cite{Note1}. 

Experimentally, \cref{eta}, $\eta/\eta_0$ starts decreasing noticeably for $\eta_0 \gtrsim 0.1$. We conclude that collision-induced scattering from surface into bulk becomes important when the surface coverage reaches $\simeq 10\%$. Taking $a \simeq 0.5\um$, our fitted value of $b \simeq k R / a = 0.7 \pm 0.1$ gives $k \simeq 0.02 \pm 0.003$, i.e. there is a $\sim 2\percent$ probability of cell-cell scattering leading to a cell leaving the surface. This parameter is difficult to determine directly from observations, but visual inspection of Supplementary Movie~1 \cite{Note1} suggests that $k \simeq 2\%$ is not unreasonable.  

There are apparent similarities between our kinetic model, \cref{model}, and one proposed recently \cite{Redner} to explain activity-driven phase separation \cite{Yang}. However, the absence of particle-particle interaction in \cite{Yang,Redner} gives rise to coexistence been liquid and vapor phases each of fixed density. On the other hand, the density dependence of the fluxes in \cref{model} means that our surface layer and bulk densities change with the average cell concentration, and what we observed is incompatible with the coexistence of two phases with invariant densities.  

We next simulated $N_0$ spherocylinders (end-to-end $\ell = 2\um$, diameter $0.8\um$) initially distributed uniformly inside a rigid sphere of radius $R$. Each `cell' self propels at $v = 20\umpers$ along its long axis \footnote{We have confirmed that using a more exact `match' with experiments, i.e. $v = 13\mu$m/s, does not change any of our conclusions.} and diffuses translationally and rotationally with isotropic diffusivities $D = 0.2\umsqpers$ and $\mathrm{D_r} = 0.2\pers$ respectively, mimicking \textit{E.~coli}. Surface cells can reorient away from the surface with rotational diffusivity $\mathrm{D_r}$. 
What happens when two motile {\it E. coli} cells collide is likely very dependent on details \cite{GoldsteinWall}. To access the essential physics, we simulated this process using a single phenomenological parameter, $\theta_r^{(c)}$, the `angular deflection at collision': two colliding cells (bulk or surface) change their propulsion directions by an angle chosen uniformly from $\left[0,\theta_\mathrm{r}^\mathrm{(c)}\right]$, and rotates around the original direction by an angle chosen uniformly in $[-\pi,\pi]$. We used 0.1~ms time steps \cite{Note1}.

If $\theta_r^{(c)} = 0$, two colliding cells remain `stuck' until Brownian motion `frees' them. Since our data suggest that collision-induced movement of cells away from the surface layer is important, we explore first the opposite limit of maximal reorientation, viz., $\theta_r^{(c)} = \pi$, which reproduces the most prominent aspects of the observed phenomenology, \cref{histograms}: a peak in $\rho(r)/\rho_0$ that decreases as $\phi_0$ increases, with a uniformly increasing bulk density.

The peak width, $\Delta$, is narrower in simulations than in experiments. In the former, $\Delta \simeq 0.05R \simeq 1 \um$, about half of the simulated cell length, $\ell = 2 \um$. Previous simulations of `wall hugging' at moderate propulsion forces \cite{Gompper} have also found $\Delta \approx \ell/2$, which is explained as a remnant of the depletion zone next to a wall in the case of passive hard rods. The wider experimental peaks, $\Delta \simeq 3 \um$, presumably is partly because real {\it E. coli} with flagella behave as rods considerably longer than $\ell = 2 \um$. However, already-noted near-edge optical aberrations may also contribute to the apparent $\Delta$.

We turn next to the number of cells in the surface layer. As expected, the case of no collision-induced reorientation, $\theta_\mathrm{r}^\mathrm{(c)}  = 0$, gives a constant $\eta/\eta_0$ as $\eta_0$ increases, \cref{eta} ({\color{blue} $\blacktriangledown$}). This does not reproduce our data. Complete randomisation at collision, $\theta_\mathrm{r}^\mathrm{(c)} = \pi$, \cref{eta} ({\color{green} $\ast$}), is more realistic. The actual trend, lying between these two limits, is account for by $\theta_r^{(c)} \simeq 0.032$, \cref{eta} ({\color{cyan} $\blacktriangle$}). To make sense of this value, note that the average reorientation at a cell-cell collision given by a particular value of $\theta_r^{(c)}$ can be recast as an effective collisional rotational diffusivity, $\mathrm{D_r^{(c)}} = \theta_r^{(c)2}/6\Delta t$, where $\Delta t = 0.1$~ms is our time step \cite{Note1}. Thus, our data suggest $\mathrm{D_r^{(c)}} \simeq 1 \mathrm{s}^{-1}$, which is about five times the Brownian $\mathrm{D_r}$. The plots of $\rho(r)/\rho_0$ at various cell densities at this value of $\theta_r^{(c)}$ (Fig.~S5) display the same phenomenology as those shown in \cref{histograms}, although the uniform rise of the bulk density with $\phi_0$ is not as rapid as observed. 

It would be unrealistic to expect our simple simulated model to reproduce exactly the totality of the data shown in Figs.~\ref{histograms} and \ref{eta}. Most importantly, the details of surface swimming depends sensitively on precise geometric parameters of the swimmers~\cite{Smith2010}. Furthermore, the distance between a wall-hugging cell and the surface can fluctuate by up to a cell width or more \cite{FordTracking,TangPNAS}, partly due to the complicated `wobble' of the cell body; and the effect of surface curvature remains largely unexplored. A basic model in which cells arriving at a surface simply align perfectly with it cannot be expected to account for such complexities, and therefore of the shape of the surface peak, \cref{histograms}. Such complexities may have less effect on an `integral measure' such as the total number of trapped cells, \cref{eta}, which is indeed what we have fitted to theory and simulations.



Our system shows certain similarities with a confined classical rarified gas in which the mean free path, $\lambda$, is larger than  or comparable to the confinement length, $L$, i.e.~the Knudsen number, $\mathrm{Kn} = \lambda / L > 1$. In both cases the particles can traverse the confined space in a straight line. For hard spheres of radius $a$ at volume fraction $\phi_0$ confined to a sphere of radius $R$, $\mathrm{Kn} = 1 / (6 \sqrt{2} \phi_0 R / a) $~\cite{Cieplak1999}, which for the droplets reported in \cref{histograms} ranges from $\mathrm{Kn} \simeq 1.9$ at $\phi_0 = 0.22\percent$ through $\mathrm{Kn} \simeq 0.5$ at $\phi_0 = 0.79\percent$ to $\mathrm{Kn} \simeq 0.06$ at $\phi_0 = 6.2\percent$. Simulations~\cite{Cieplak1999} show that in a confined rarefied gas at $\mathrm{Kn} > 1$ with attractive walls, the evolution of the density profile as a function of average gas density is closely similar to that shown in \cref{histograms}: a broadening surface peak and uniformly increasing bulk density. The `attraction' in our case comes from `wall hugging'~\cite{li2009accumulation,berke2008hydrodynamic}. 

\begin{figure}
    \centering

    \includegraphics[width = 0.7\columnwidth]{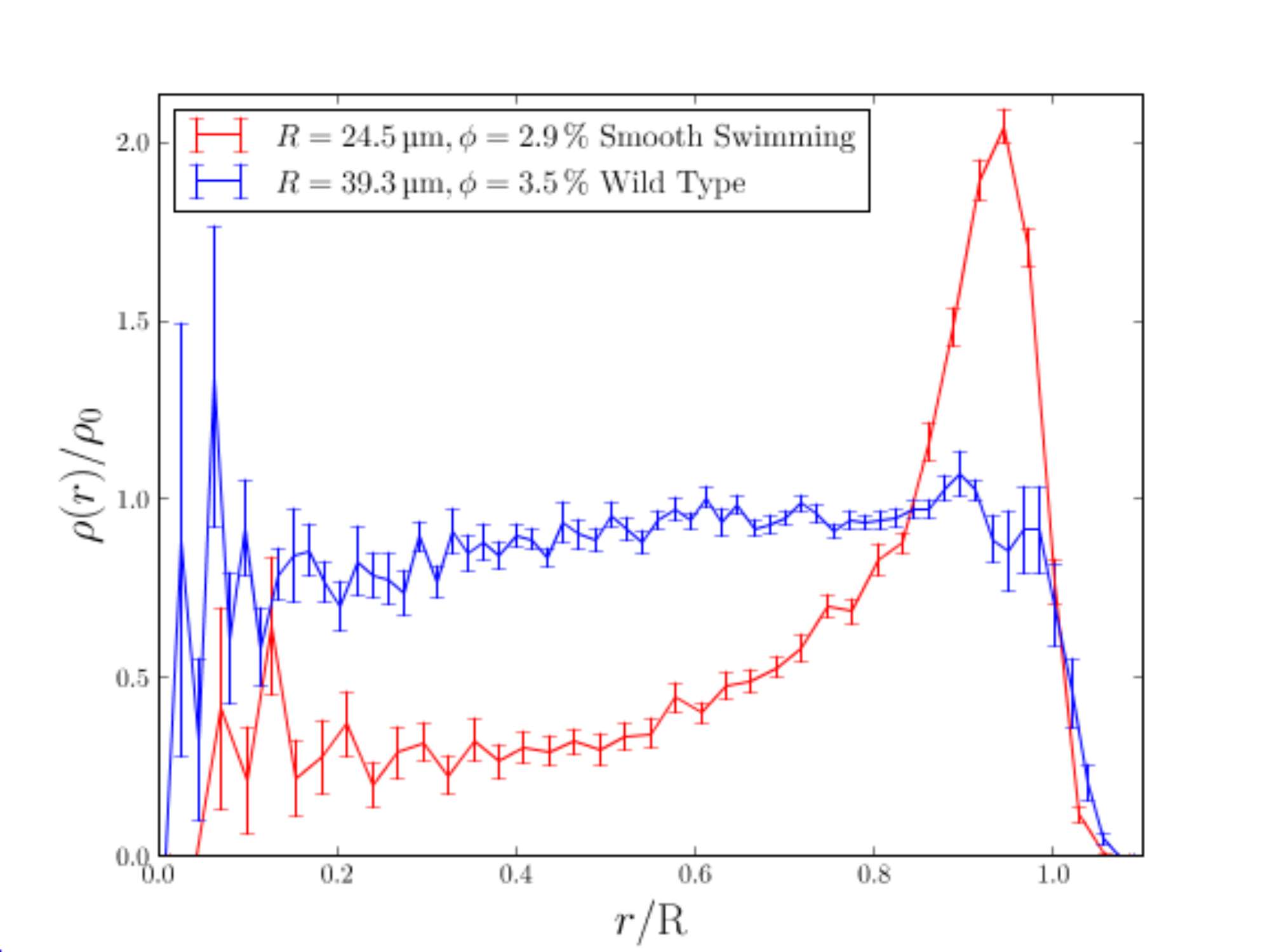}

    \caption{Comparison of density profiles for smooth swimmers (red) and wild-type run-and-tumble cells (blue).}

    \label{wild_type}
\end{figure}

This analogy is no longer appropriate either at large $\phi_0$ or when the persistence length of the swimmers drops below the system size, $\lambda / 2 R < 1$. The latter can be probed using wild-type cells, which tumble every $1\second$ or so between straight `runs'. Given $v \lesssim 20\umpers$, we now have $\lambda \lesssim 20\um$. \cref{wild_type} compares the density profile for a smooth swimmer in a drop with $2R \ll \lambda$, and a wild type in a drop with $2R > \lambda$. The density peak at the droplet edge has disappeared in the latter case, presumably because surface tumbles now remove cells from the trapped layer too rapidly for a peak to build up.

At cell densities 10 times higher than the maximum reported so far, we observed vortices with constantly changing orientations inside our emulsion drops (Supplementary Movie~3 \cite{Note1}). This may be compared to {\it B. subtilise} confined to cylindrical water drops at high densities, where a single vortex aligned to the cylinder axis is seen~\cite{wioland2013confinement}. The difference may partly be due to the greater length of {\it B. subtilis} cells, and partly to differing spatial symmetry. Such collective motion is left to future work. 

Finally, if internal flows from bacterial motility can set up exterior flows, then our droplets should display at least enhanced positional fluctuations. Tracking revealed no such activity. This is likely because the lecithin layer stabilising each droplet is rigid enough to decouple internal and external flows. 

To summarise, we have observed the emergence of many-body behavior in spherical water droplets filled with increasing density of motile {\it E. coli} bacteria. The single-body physics of previously studied `wall hugging'~\cite{li2009accumulation,berke2008hydrodynamic} together with Brownian reorientation taking cells from the surface into the bulk suffice to explain observations up to a surface coverage of $\eta_0 \simeq 0.1$, \cref{eta}. Thereafter, the decrease of $\eta/\eta_0$ with $\eta_0$ evidences cell-cell scattering; fitting to a simple theory suggests that a few per cent of collisions scatter cells from the surface layer to the bulk. Bulk `traffic' of cells from one part of the inner surface to another, initially due solely to Brownian reorientation, and then increasingly due to cell-cell scattering, explains the observe uniform increase in the bulk density as the average cell density increases.

\begin{acknowledgments}
We thank A. Brown, A. Dawson, D. Dell'Arciprete, A. Jepson, and T. Pilizota, for discussions. The work was funded by the Royal Society, the UK Engineering and Physical Sciences Research Council (EP/I004262/1, EP/J007404/1), the European Union (FP7-PEOPLE (PIIF-GA-2010-276190)) and the European Research Council (ADG-PHYAPS).
\end{acknowledgments}

\bibliography{bacteria_in_drops}

\newpage

\begin{center}
{\Large Supporting Information}
\end{center}

\setcounter{footnote}{0}
\setcounter{equation}{0}
\setcounter{figure}{0}

\renewcommand{\thefigure}{S\arabic{figure}}
\renewcommand{\theequation}{S\arabic{equation}}

\subsection{Sample preparation}

P1 phage transduction~\cite{miller1972experiments} was used to create a smooth swimming strain (AB1157 $\Delta cheY$) using the appropriate \textit{E. coli} K12 single knockout mutant from the KEIO collection~\cite{baba2006construction}. Kanamycin (final concentration $30\ugperml$) was added to all growth media for AB1157 $\Delta cheY$. The GFP encoding plasmid pHC60 was extracted using a QIAGEN Plasmid mini kit and transformed into AB1157 $\Delta cheY$ using a method based on CaCl$_2$ as detailed in~\cite{sambrook2001molecular}. Tetracycline (final concentration $5\ugperml$) was added to maintain pHC60.

Bacteria were grown overnight in Luria-Bertani broth at $30\degree$C shaken at 200 rpm; harvested in the exponential phase; washed three times by careful filtration with a $0.45\um$ filter and resuspended in a phosphate motility buffer ($6.2\mM$ \ce{K_2HPO_4}; $3.8\mM$ \ce{KH_2PO_4}; $67\mM$ \ce{NaCl}; $0.1\mM$ \ce{EDTA} at $7.0\pH$) to optical densities of OD = 1~to~3 (at $600\nm$), corresponding (from plate count) to cell densities of $1.55 \times 10^9\perml$ to $4.65 \times 10^9\perml$.

Emulsions were obtained by mechanically dispersing a small amount of bacterial suspension ($\sim 2\percent$v/v) in sunflower oil (Sigma, used as purchased). We could create droplets stable for days without adding surfactant, presumably due to native lecithins in the oil~\cite{grompone2004sunflower}. Samples of emulsion ($\sim 400 \,\upmu$l) were loaded into $8 \times 8 \times 8\mm^3$ open coverglass chambers for microscopic observation from below.

\subsection{In situ characterisation}

\textit{In situ} differential dynamic microscopy (DDM)~\cite{wilson2011differential,martinez2012differential} was performed to measure the swimming speed distribution, $P(v)$, of the cells within individual droplets and the fraction of non-motile organisms. We found an average swimming speed inside a droplet to be $\bar{v} \simeq 13.5 \pm 0.7 \umpers$ at cell concentrations similar to those used in our experiment (OD=2, initial cell density used for emulsification).  The fraction of non-motile organisms inside an emulsion drop was around $30\percent$. Interestingly, at high cell densities (OD=24, initial cell density used for emulsification) although the average swimming speed was $\bar{v} \simeq 12\umpers$, the fraction of non-motile organisms decreases to around $10\percent$.


The oxygenation condition inside droplets was monitored using a ruthenium dye, RTDP~\cite{RTDP}, whose fluorescence is quenched by molecular oxygen. We found using DDM that the dye did not affect cell motility at concentrations $\lesssim 50\uM$.  Imaging the intensity of RTDP-containing material droplets showed that our cells experienced spatio-temporally constant oxygen conditions for many hours. In particular, there were no  oxygen gradients near droplet edges, so that oxytaxis is absent. Presumably, as in~\cite{wioland2013confinement}, the high solubility of oxygen in oil~\cite{battino1981oxygen} keeps each drop well oxygenated.

\subsection{Imaging \& image reconstruction}

We studied the spatial distribution of cells in bacteria-containing droplets using confocal microscopy. A green channel imaged GFP-labelled cells. The dye 1,1$^\prime$-Dioctadecyl-3,3,3$^\prime$,3$^\prime$-Tetramethylindocarbo-cyanine perchlorate (Dil, Molecular Probes) was dissolved in the oil to highlight in a red channel the droplet-stabilising lecithins, with which it has an enhanced affinity. Each image, taken in an inverted Zeiss AXIO Observer.Z1 microscope with an LSM700 scanning module and a $60\times$ oil immersion objective, covered a $193\um \times 193\um$ field to an optical depth of $2\um$, and typically showed $\lesssim 10$ droplets with various $(R,\rho_0)$. We acquired $z$-stacks of typically 30 images (256 lines, $0.8\second$ per slice) spaced at $2\um$ of all the droplets in a field of view, starting several $\um$ below the bottom of the sample compartment (Supplementary Movie 1). All images were acquired within 2 hours of sample preparation.

Local intensity maxima were identified in the green channel (GFP-labelled bacteria, excited at 488~nm) of Fiji-filtered~\cite{Fiji} images to give cell coordinates, while red channel (Dil-stained oil, excited at 555~nm) images were analysed to yield droplet centres and radii (see Fig.~1 in main text).  The droplet were approximately spherical. We select typically 12 points at the w/o interface of each droplet within its stack of images and fit these points to an ellipsoid~\cite{bonej}. We found $0.03 \leq 1 - c / a \leq 0.17$, where $a$ and $c < a$ are the longest and shortest principal axes, with little systematic correlation between $c/a$ and droplet size. 

\begin{figure}[h]
    \centering

    \includegraphics[width = 0.5 \textwidth]{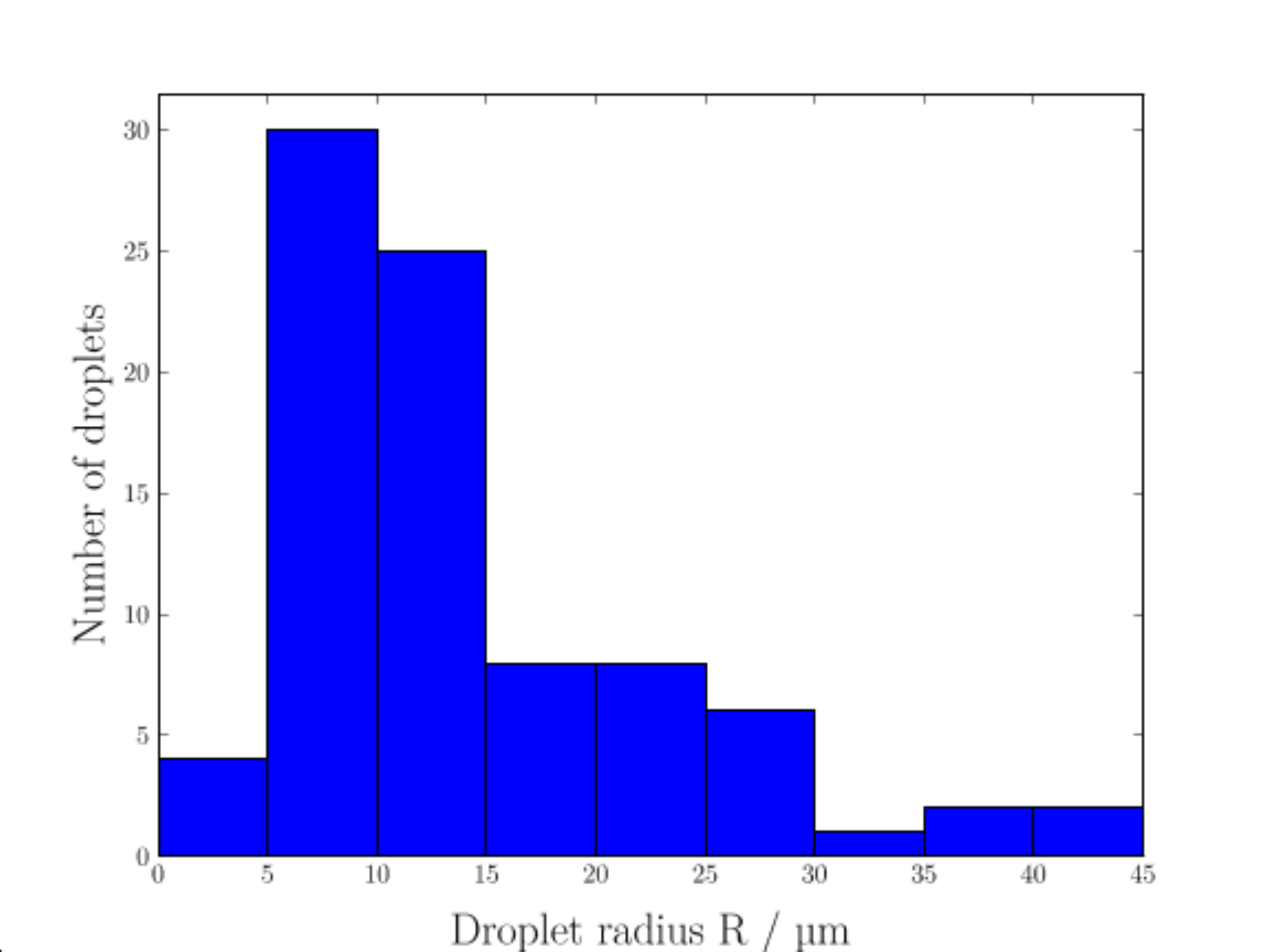}

    \caption{Distribution of droplet radii in our emulsions.} 

    \label{emulsioncharacterization}
\end{figure}

The measured distribution of average droplet radius, $R$, is shown in \cref{emulsioncharacterization}.

{\bf Supplemental Movie 1}: Confocal $z$-stack of a typical w/o emulsion that encapsulates active bacteria. a) Processed stack. b) Overlay of the positions of bacteria with the unprocessed stack.

{\bf Supplemental Movie 2}: Phase-contrast movie of a typical w/o emulsion that encapsulates active bacteria.

\subsection{Optical Distortions}

In principle, counting cells within concentric shells (`bins') of width $\Delta r$ gives the cell density as a function of distance from the center, $\rho(r)$, which we assume to be isotropic. However, it is well known that refractive index differences within a sample induce distortions in the image~\cite{Schwertner2007}. These distortions can range from simple local blurring of features (\emph{i.e.} reduction in resolution) to severe 3-dimensional geometric distortions, producing images with little resemblence to the object. In the current context, bacteria are imaged through a smooth spherical interface between the sunflower oil ($\mathrm{n_o}=1.46$) and the aqueous buffer ($\mathrm{n_w}=1.33$). 

In order to assess the nature and severity of the resultant distortions, we modelled confocal imaging of objects within water droplet using a commercial ray tracing package (ZEMAX 13 Professional, Radiant Zemax, LLC). The modelled microscope is based around a publicly available desription of a high numerical aperture oil immersion lens~\cite{laikin2010lens}. It assumes illumination using a blue light point source ($\lambda = 450\nm$) and confocal detection, with signal from either green fluorescent objects or red fluorescent oil (with increased fluorescence in a thin shell surrounding the water droplet) being recorded. For simplicity, transverse scanning was simulated by moving the sample through a fixed laser focus rather than scanning the laser as in the experiments. Axial scans were performed by changing the thickness of the oil immersion layer by a prescribed amount $\dif z$.

\begin{figure}
    \centering

    \includegraphics[width = 0.4 \textwidth]{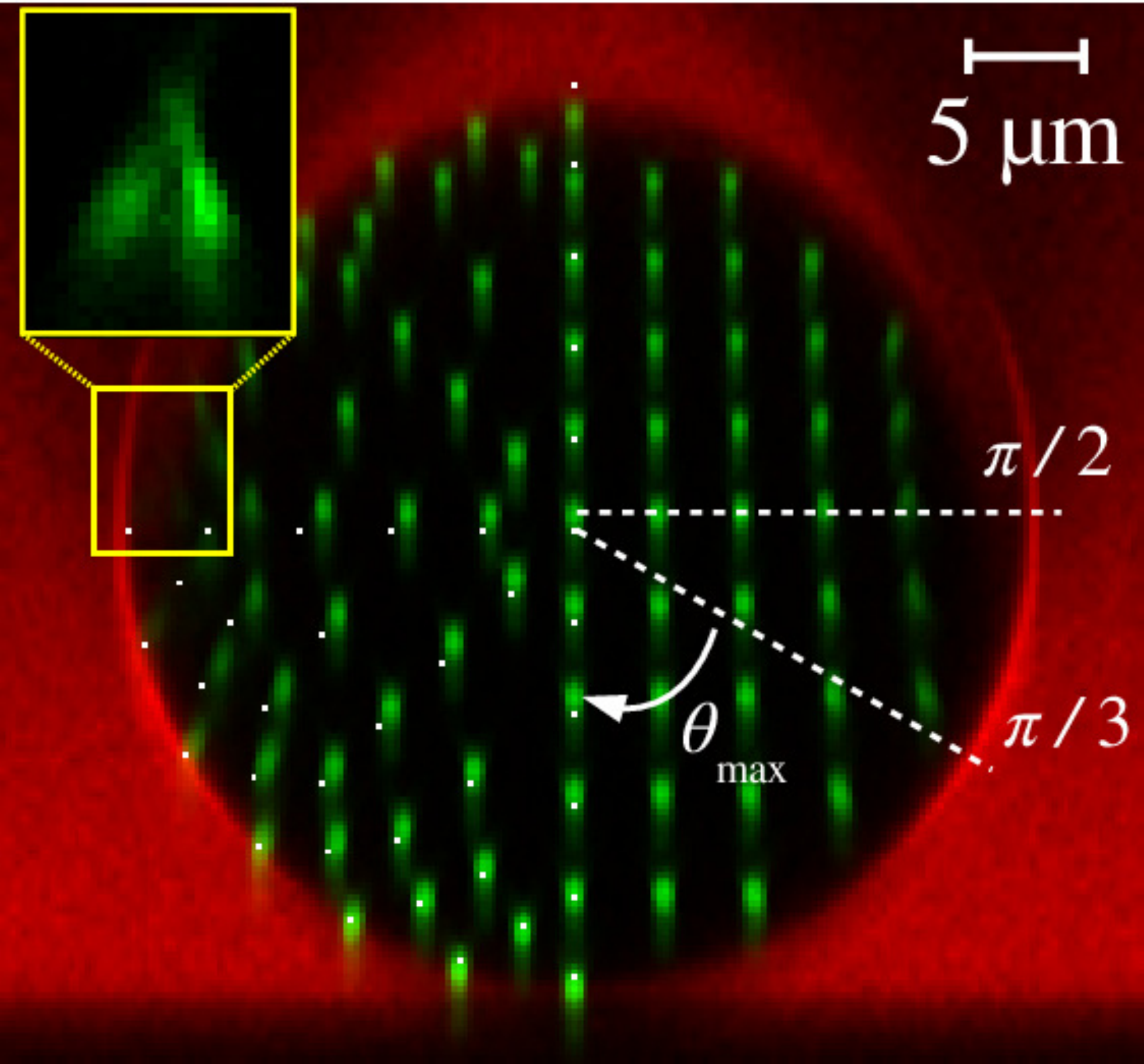}
    
    \caption{Simulation of optical distortion due to imaging inside a spherical droplet of higher refractive index than the surrounding medium. The outline of the upper hemisphere is severely distorted. White points inside become blurred and shifted into the green patches. Positional errors increase as the angle from the central axis increases, to the extent that (inset) an object on the equator can become two. Thus, we only used points within a $\pi/3$ cone to acquire quantity data for calculating $\rho(r)$.}
    
    \label{optics}
\end{figure}

Representative results showing an $x-z$ cross-section passing through the centre of a droplet of radius $20\um$ are presented in \cref{optics}. Distortions within the upper hemisphere are evident, both in the red signal from the oil, which gives an upper hemisphere with an apparently smaller radius, and the location of the green objects. Therefore only experimental data from the lower hemisphere was included in our analysis. However, even within the lower hemisphere distortions are noticeable.

Within a cylinder of radius $\sim 0.75R$ the shape and contrast of the green objects are hardly affected and the position shows a relatively small systematic underestimation of the axial distance ($\sim 10\percent$). This implies in particular that the shape of the radial bacteria density distribution near the droplet centre is not affected by imaging distortions. However, for large transverse distances the contrast and shape are also clearly affected, leading to a low contrast region near the equator. The distortions of the particle shape are highlighted by a sphere touching the droplet surface at the equator (see inset): it shows up as 2 separate, almost distinct maxima. The onset of this region of low contrast and severe distortions can be roughly estimated by looking at a ray in $z$-direction hitting the spherical interface at a distance $x$ from the central axis of the droplet. Its angle of incidence is $\sin(\theta) = x / R$, and within the droplet Snell's law gives $\sin(\theta')= n_o / n_w \sin(\theta) = (n_o/n_w) (x / R)$. However, beyond the critical angle $\theta_c = \sin^{-1}(n_w / n_o) \approx 66\degree$ the ray can not actually enter the droplets any more, giving rise to the `blind' region. It was therefore decided to restrict analysis of the experimental data to the bottom sector with $\theta_S = 60\degree$ opening angle.

\begin{figure}
    \centering

    \includegraphics[width = 0.45 \textwidth]{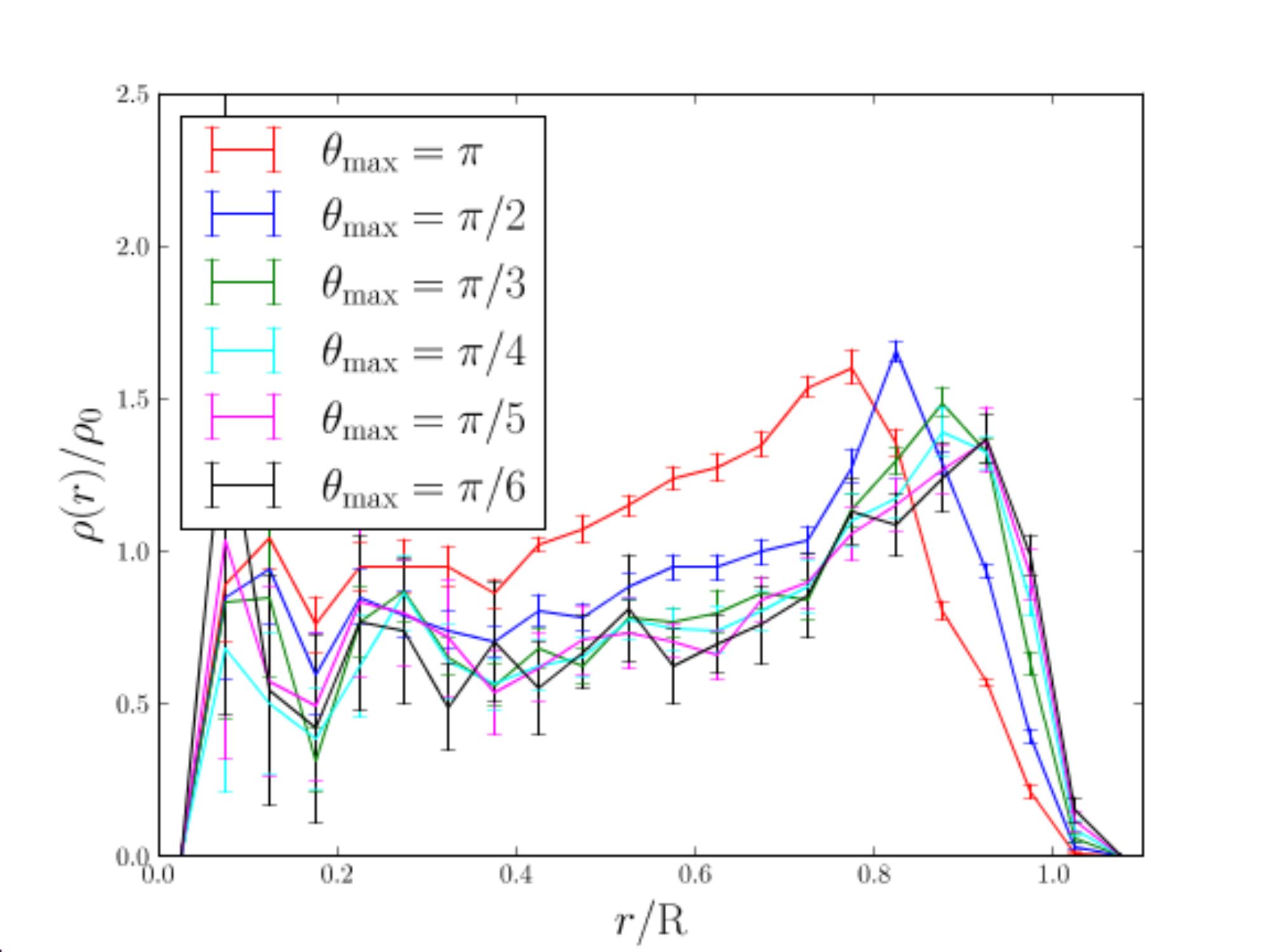}

    \caption{Comparison of the bacterial radial number density distribution for one droplet, considering measured bacterial positions only within a spherical cone of half-angle $\theta_\mathrm{max}$ from the imaging axis. At large deviations from the imaging axis data points are distorted by spherical abberations. As $\theta_\mathrm{max}$ is decreased, the distribution converges on a robust shape. In our data analysis we chose $\theta_\mathrm{max}=\pi/3$, as a compromise between minimising random and systematic (optical distortive) errors.} \label{aberration}

    \label{angles}
\end{figure}

\begin{figure}
    \centering

    \includegraphics[width = 0.45 \textwidth]{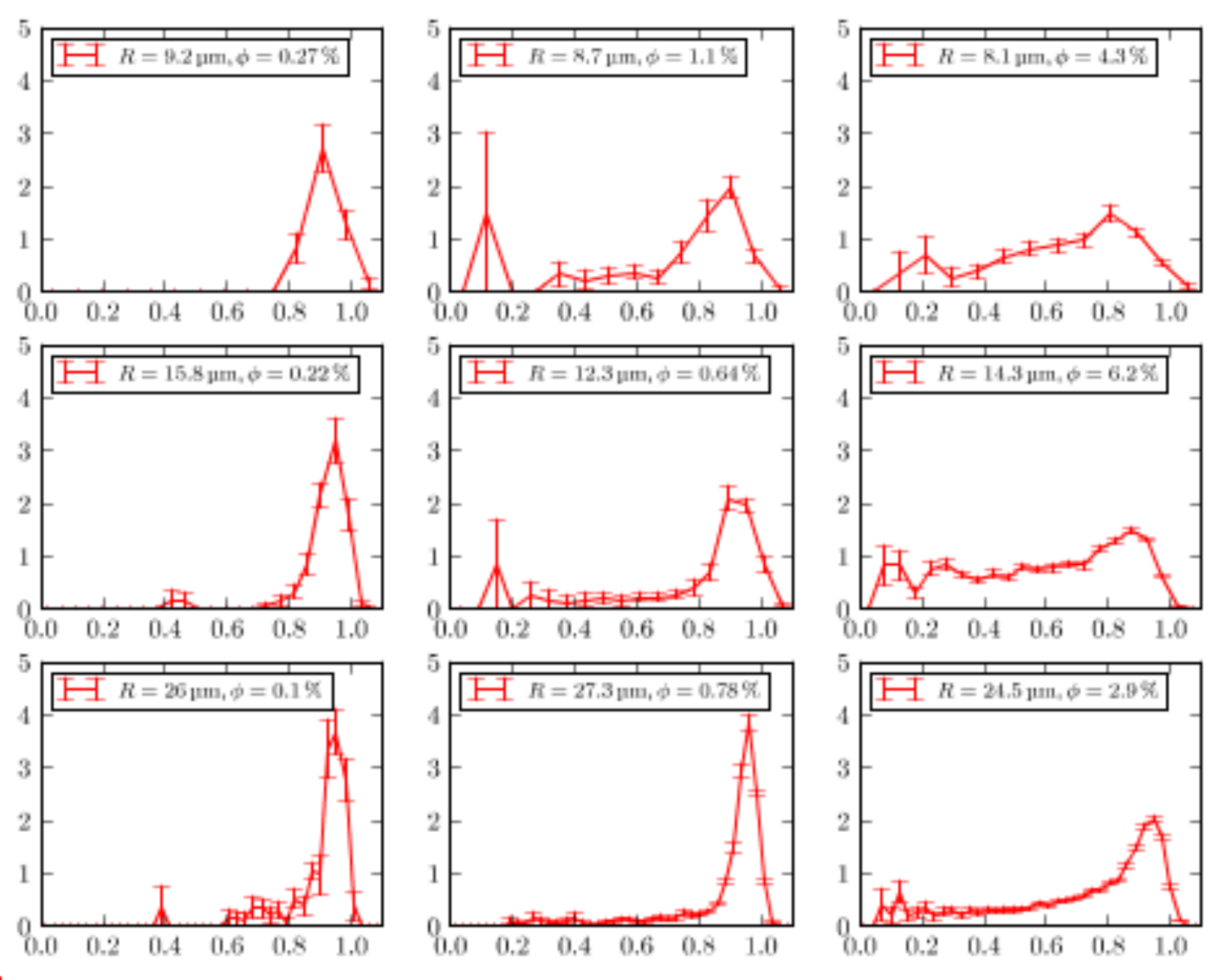}
    \includegraphics[width = 0.45 \textwidth]{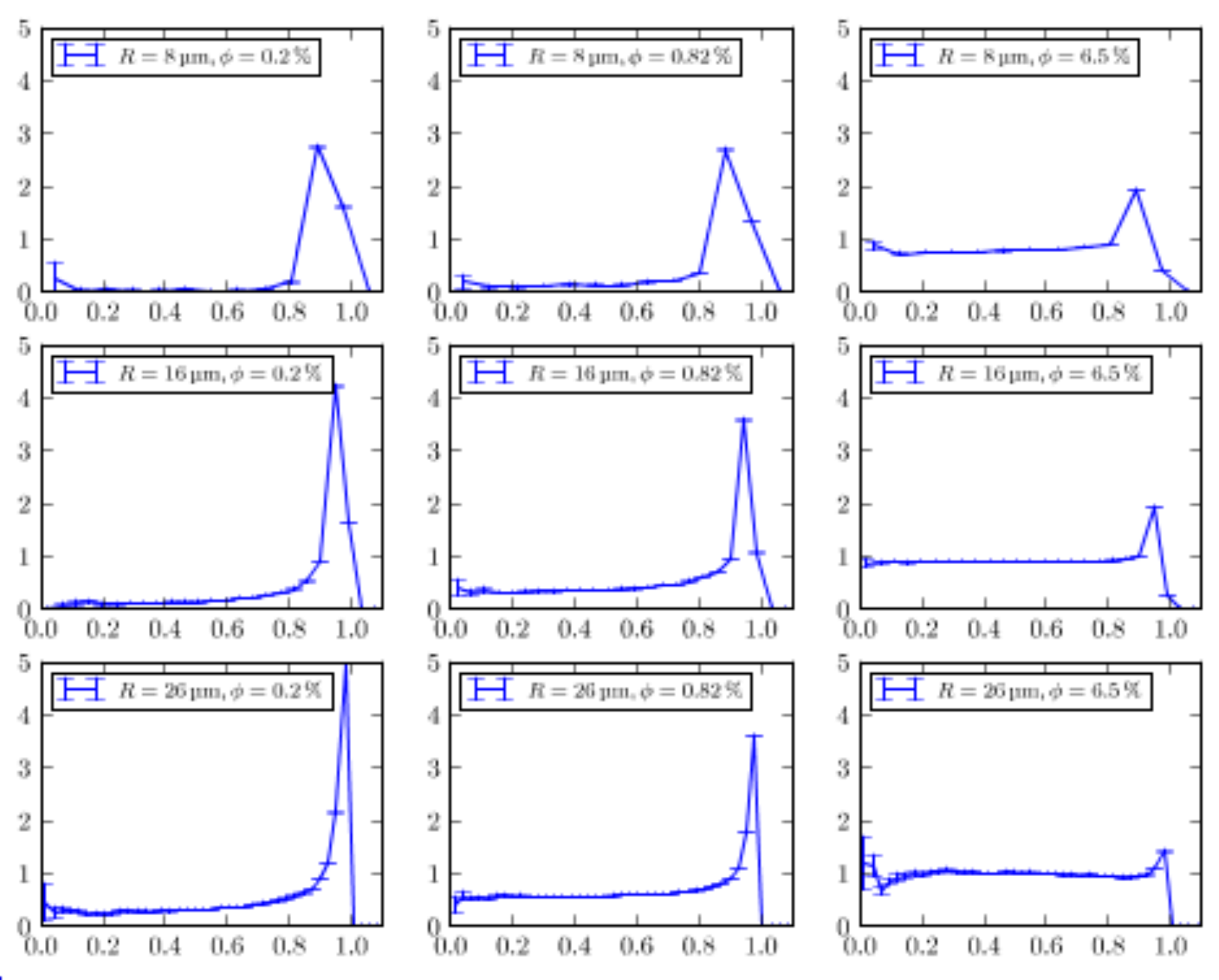}
    
    \caption{Radial bacterial number density distributions, $\rho(r)$, normalised by the average number density of the whole droplet, $\rho_0$ and the droplet radius, $R$, for several datasets with different radii and cell densities. Upper (red): Experiment. Lower (blue): Simulation.}

    \label{grid}
\end{figure}

Our observed shift and broadening of the peak could at least partially be due to these distortions, so the detailed shape of the radial profile plots close to the edge should not be overinterpreted, \cref{angles}. On the other hand, there appears no systematic radius dependence of the observed phenomenology, \cref{grid}.

\subsection{Diffusivities and persistence length}

We worked mostly with smooth (non-tumbling) swimmers, whose persistence length, $\lambda$, is due to loss of orientation as a result of rotational Brownian motion. The mean-squared angular drift of each cell is given by $\langle \theta^2 \rangle \simeq \mathrm{D_r} t$, where the rotational diffusivity $\mathrm{D_r}$ is controlled mainly by the length of the cell body plus flagellum. Modelling cell+flagella by an ellipsoid with semi-axes $a \simeq 5\um$ and $b \simeq 0.5\um$,  then gives $\mathrm{D_r} = k_BT (16 \pi \eta ab^2 / 3)^{-1} \simeq 0.2\pers$~\cite{berg1993random}. (This value is the same order of magnitude as but bigger than that measured recently from a related strain of {\it E. coli} \cite{wioland2013confinement}.) The rotational relaxation time is $\mathrm{D_r}^{-1} \simeq 5$~s, and the persistence length for $\bar v \simeq 20\umpers$ swimmers is $\lambda \simeq \bar v \tau_r \simeq 100\um$. Thus, all of our emulsion droplets are considerably smaller than the persistence length of our smooth swimmers. 

We next estimate upper and lower bounds of the translational (center of mass) diffusivity of the cells. If we model a cell as a sphere with volume appropriate to a $(1\um \times 2\um)$ spherocylinder, then the Stokes-Einstein relation gives $D \simeq 0.3\umsqpers$. On the other hand, modelling the cell+flagella as an ellipsoid with dimensions given above gives $D \simeq 0.1\umsqpers$. We use $D = 0.2\umsqpers$ in our simulations.

\subsection{Analytic model}

We consider a slab of bacterial solution of size $R$ with $N_b$ \emph{uniformly} distributed bacteria. Bacteria swim with a constant velocity $v$. The number of bacteria hitting the surface in time $dt$ is then $N_b v dt / R$. The probability of staying at the surface after hitting it is taken to be equal to the fraction of the surface that is free from bacteria, $1 - N_s A_b / A$, where $N_s$ is the number of bacteria at the surface, $A_b$ is the surface area covered by one bacterium and $A$ is the total surface area. Therefore, the total number of bacteria arriving and staying at the surface during time $dt$ is
\begin{equation}
    \dif N_s = \del{1 - \frac{N_s A_b}{A}} N_b \frac{v \dif t}{R} \,.
\end{equation}

We assume that there are two mechanisms for bacteria leaving the surface. First, each cell may leave by its own reorientation, with characteristic time $\tau$, \textit{i.e.} with rate $N_s (\gamma dt)$ where $\gamma = \tau^{-1}$. Secondly, a call may encounter another cell in a two-body `scattering' event, which can be modelled by $\beta N_s^2 dt$. The change in surface cell number in time $dt$ is therefore:
\begin{equation}
    \dif N_s = \del{1 - \frac{N_s A_b}{A}} N_b \frac{v \dif t}{R} - \gamma N_s \dif t - \beta N_s^2 \dif t \,,
\label{model_flux}
\end{equation}
where $\beta$ is the scattering frequency, and $\gamma$ the self-scattering frequency. These values are difficult to estimate for arbitrary surface coverages but can be calculated for low $N_s$.

With regards to inter-bacterial scattering, consider bacteria swimming at the surface and let us select one bacterium as a `target'. The probability of another bacterium hitting the target from a distance $\lambda$ is $2 a / 2 \pi \lambda$, where $a$ is the radius of a bacterium, which is assumed to have a circular projection on the surface. The number of bacteria hitting the target from a thin shell $\del{\lambda, \lambda + d \lambda}$ is
\begin{equation}
    \frac{2a}{2 \pi \lambda} \sbr{\pi \del{\lambda + \dif \lambda}^2 - \pi \lambda^2} \frac{N_s}{A} = 2 a \frac{N_s}{A} \dif \lambda \,.
\end{equation}
Since the target can be hit only from a circle of radius $v \dif t$, the total number of scattering events between all bacteria and a selected target is
\begin{equation}
    \int_0^{v \dif t} 2 a \frac{N_s}{A} \dif \lambda = 2 a v dt \frac{N_s}{A} \,.
\end{equation}
Since the same argument is valid for every bacterium on the surface, the total number of scattering events is proportional to
\begin{equation}
    \frac{1}{2} N_s 2 a v \dif t \frac{N_s}{A} = a v \dif t \frac{N_s^2}{A} \,,
\end{equation}
where the factor $1 / 2$ is introduced to properly account for the number of bacterial pairs. By comparing this expression with \cref{model_flux}), we identify
\begin{equation}
    \beta = k \frac{a v}{A}\,,
    \label{beta}
\end{equation}
where $k$ is a probability that one of the bacteria participating in a scattering event would come off the surface. Once again, this argument only properly works for a low surface coverage.
Using \cref{model_flux} with the approximation \cref{beta}, we obtain in the steady-state
\begin{equation}
    \del{1 - \frac{N_s A_b}{A}} N_b \frac{v}{R} - \gamma \frac{R}{v} N_s - k \frac{a v}{A} N_s^2 = 0 \,.
\end{equation}

Finally, observing that $N_s + N_b = N$, where $N$ is the total number of bacteria in the system, and introducing $\eta = N_s A_b / A$ and $\eta_0 = N A_b / A$, we obtain Eq.~(1) in the main text, where $b = k a R / A_b$ and $c = \gamma R / v$. Note that $\eta_0$ can be significantly larger than unity.

\subsection{Simulation algorithm}

We simulated our system in continuous, 3-dimensional space and discrete time. Thus, the algorithm iterated for each spherocylinder $\mathrm{i}$ at position $\bm{r}_\mathrm{i}$ self propelled with velocity $\bm{v}_\mathrm{i}$ parallel to the long axis, at each time-step of size $\Delta t$ is,
\begin{equation}
\begin{split}
    \bm{r}_\mathrm{i}(t + \Delta t) & = \bm{r}_\mathrm{i}(t) + \bm{v}_\mathrm{i}(t) \Delta t + \sqrt{2 \mathrm{D} \Delta t} \bm{\eta}_\mathrm{i}(t) \,, \\
    \bm{v}_\mathrm{i}(t + \Delta t) & = \bm{R}(\theta \eta_\mathrm{i, r}(t)) \bm{v}_\mathrm{i}(t) \,,
\end{split}
\end{equation}
where $\bm{\eta}$ is a vector of unit gaussian noise representing translational Brownian motion with diffusion constant $\mathrm{D}$ and $\bm{R}$ is a rotation matrix representing random rotational noise of magnitude $\theta_\mathrm{r}$ ($\eta_\mathrm{i, r}(t)$ is a sample from a unit gaussian distribution).

The magnitude of rotational noise $\theta$ is determined \textit{via}
\begin{equation}
    \theta_\mathrm{r} = \sqrt{2 \mathrm{D_r} \Delta t} +
    \begin{cases}
        \theta_\mathrm{r}^\mathrm{(c)} & \quad \text{collision} \\
        0                              & \quad \text{no collision}
    \end{cases}
\end{equation}
where $\mathrm{D_r}$ is the bulk rotational diffusion constant, and $\theta_\mathrm{r}^\mathrm{(c)}$ is a maximum deflection angle in $[0,\pi]$.

If two cells overlap after propagation over a time step, then the two cells are `back tracked' to their previous position. In subsequent time steps, Brownian motion ($\mathrm{D}$ and $\mathrm{D_r}$) and collisional deflection, $\theta_\mathrm{r}^\mathrm{(c)}$, eventually free the two cells. If $\theta_\mathrm{r}^\mathrm{(c)} = \pi$, then the two cells take randomised orientations during a single time step.

If a cell overlaps with a surface after propagation over a time step, the cell is `back tracked' to a position where the overlap vanishes, and the velocity is aligned parallel to the surface,
\begin{equation}
\begin{split}
        \bm{\hat{v}}_\mathrm{i \parallel}(t) & = \dfrac{\bm{v}_\mathrm{i}(t) - \del{\bm{v}_\mathrm{i}(t) \cdot \hat{\bm{u}}} \bm{v}_\mathrm{i}(t)}{\abs{\bm{v}_\mathrm{i}(t) - \del{\bm{v}_\mathrm{i}(t) \cdot \hat{\bm{u}}} \bm{v}_\mathrm{i}(t)}} \\
        \bm{v}_\mathrm{i}(t + \Delta t) & = \abs{\bm{v}_\mathrm{i}(t)} \bm{\hat{v}}_\mathrm{i \parallel}(t) \,,
\end{split}
\end{equation}
where $\hat{\bm{u}}$ is the outer-pointing normal to the droplet surface.

Finally, to understand the significance of the parameter $\theta_r^{(c)}$, we recast it in terms of an effective collision-induced diffusivity. The variance of the uniform distribution on $[-\theta_r^{(c)}, \theta_r^{(c)}]$ out of which we choose the reorientation angle at collision is $\sigma^2 = \theta_r^{(c)2}/3$. To map this to an effective rotational diffusion with diffusivity $\mathrm{D}_r^{(c)}$, we recall that this process produces a Gaussian distribution with variance $\sigma^2 = 2\mathrm{D}_r^{(c)} \Delta t$ that scales linearly with the time interval, which we take as our time step ($1 \us$). Equating variances gives the expression quoted in the main text: $\mathrm{D}_r^{(c)} = \theta_r^{(c)2}/6\Delta t$.

\begin{figure}
    \centering

    \includegraphics[width = 0.4 \textwidth]{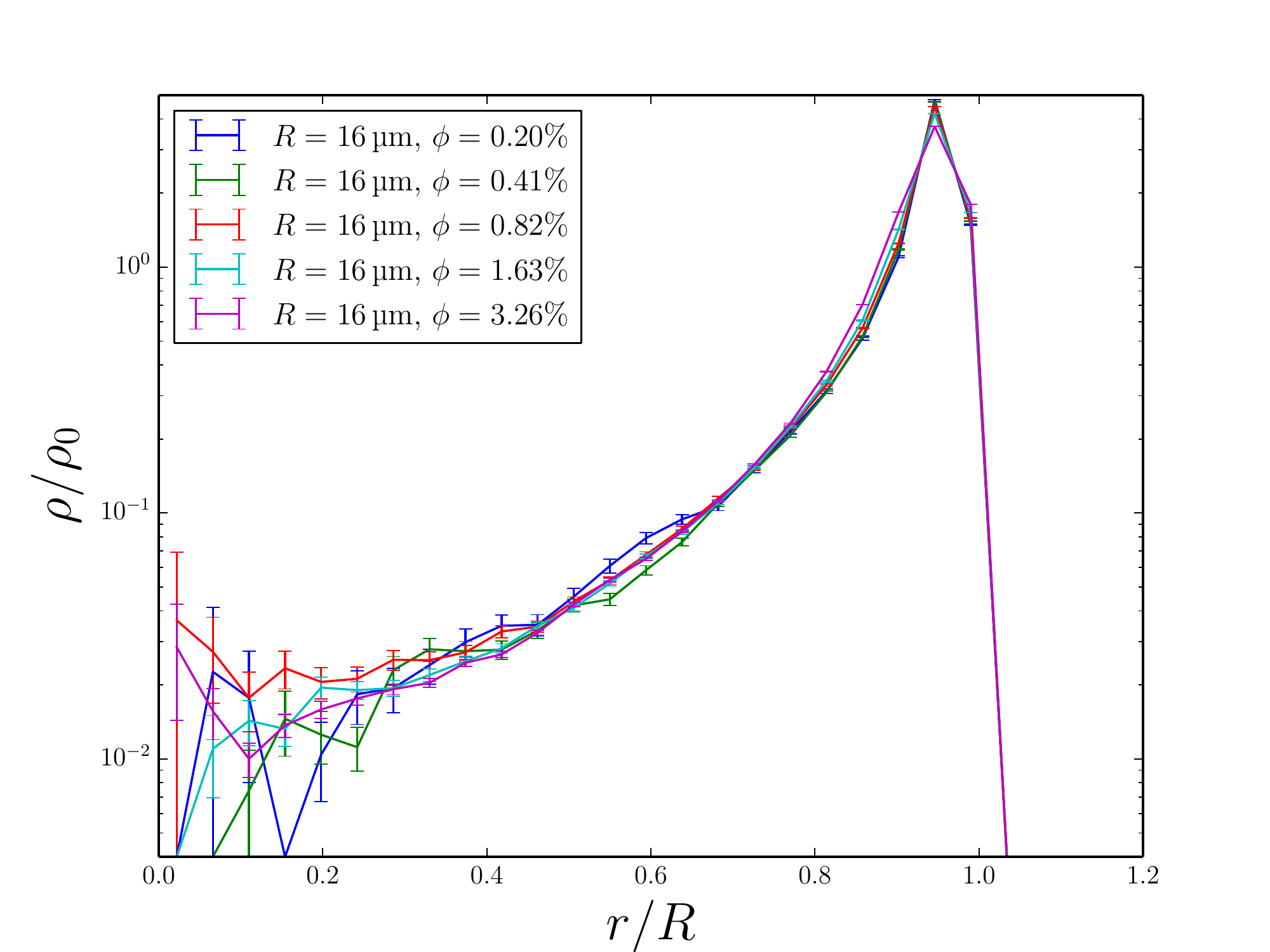}
    \includegraphics[width = 0.4 \textwidth]{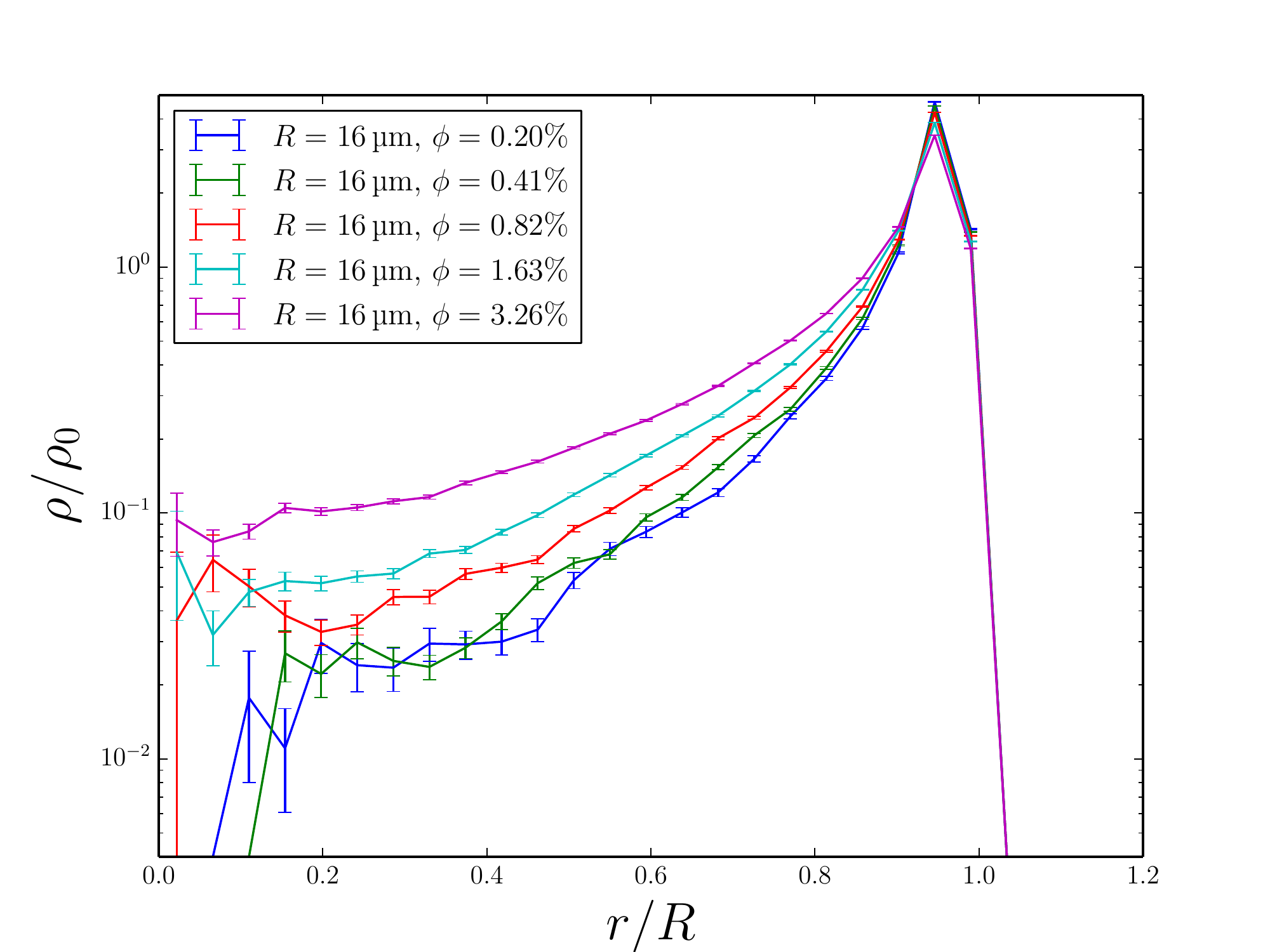}
    
    \caption{Radial bacterial number density distributions, $\rho(r)$, normalised by the average number density of the whole droplet, $\rho_0$, plotted against the radial distance from the center, $r$, normalised by the droplet radius, $R$, averaged over 10 data sets for $R = 16\um$. Upper: No collision-induced reorientation, $\theta_r^{(c)} = 0$. Lower: Intermediate level of collision-induced reorientation, $\theta_r^{(c)} = 0.032$. We have used a log vertical scale to highlight the fart that the bulk density does {\it not} rise in the case of $\theta_r^{(c)} = 0$, and the rise is less rapid than is seen in either plots in Fig.~2 of the main text in the case of $\theta_r^{(c)} = 0.032$.}
    \label{fitted_peaks}
\end{figure}

\subsection{Surface peak evolution with fitted value of $\theta_r^{(c)}$}

Figure~\ref{fitted_peaks} shows how the surface peak evolves with average cell volume fraction, $\phi_0$ using the value of $\theta_r^{(c)} = 0.032$, which corresponds to a collision-induced rotational diffusivity that is about 5 times the Brownian rotational diffusivity. A plot of comparable experimental results (those shown in Fig. 2a of the main text) is also shown. These simulations give a very similar phenomenology, but the quantitative comparison with experiments is less exact than the case of full randomisation at collisions, especially at the highest $\phi_0$ shown. \\

\subsection{Collective motion}

We show in a movie the emergence of collective vortices in droplets with cell densities an order of magnitude higher than the highest $\phi_0$ considered in the main text, with an estimated $\phi_0 \sim 30\%$. 

{\bf Supplemental Movie 3}: Phase-contrast movie of a an emulsion drop encapsulating swimming {\it E. coli} at a density of $\phi_0$ estimated at 30\%. Note the existence of large-scale ($\sim$ droplet size) vortices. 

\bibliography{bacteria_in_drops}


\end{document}